\pdfoutput=1

\documentclass[aps,pra,10pt,twocolumn,showpacs,superscriptaddress,nofootinbib]{revtex4}
\usepackage{amsmath}
\usepackage{latexsym}
\usepackage{amssymb}
\usepackage{graphics,epstopdf}

\begin{document}

\title{Einstein-Podolsky-Rosen-steering using quantum correlations in non-Gaussian entangled states}

\author{Priyanka Chowdhury}

\affiliation{S. N. Bose National Centre for Basic Sciences, Block JD, Sector III, Salt Lake, Kolkata 700098, India}

\author{Tanumoy Pramanik}

\affiliation{S. N. Bose National Centre for Basic Sciences, Block JD, Sector III, Salt Lake, Kolkata 700098, India}

\author{A. S. Majumdar}

\affiliation{S. N. Bose National Centre for Basic Sciences, Block JD, Sector III, Salt Lake, Kolkata 700098, India}

\author{G. S. Agarwal}

\affiliation{Department of Physics, Oklahoma State University, Stillwater, Oklahoma 74078, USA}

\begin{abstract}
In view of the increasing importance of non-Gaussian entangled states in
quantum information protocols like teleportation and violations of Bell
inequalities, the steering of continuous variable non-Gaussian
entangled states is investigated. The EPR steering for Gaussian states may be 
demonstrated through 
the violation of the Reid inequality involving products of the inferred
variances of 
non-commuting observables. However, for arbitrary states the Reid inequality
is not always necessary because of the higher order 
correlations in such  states. One then needs to use  the entropic steering 
inequality. 
We examine several classes of currently important 
non-Gaussian entangled states, such as the 
two-dimensional harmonic 
oscillator, the photon subtracted two mode squeezed vacuum, and the NOON state,
in order to demonstrate the steering property of such states. A comparative 
study of the violation of the Bell-inequality 
for these states shows that the entanglement present is more easily revealed 
through steering compared to Bell-violation for several such states. 


\end{abstract}

\date{\today}

\maketitle

\section{Introduction}

The pioneering work of Einstein, Podolsky and Rosen (EPR) \cite{epr} has over 
the years lead to the unfolding of several rich and arguably  paradoxical
features of quantum mechanics \cite{Reid_09}. Considering a position-momentum correlated state
of two particles, and assuming the notions of spatial separability, locality,
and reality to hold true at the level of quantum particles, EPR argued that
the quantum mechanical description of the state of a particle is not
complete. An immediate consequence of correlations between spatially separated
particles was then noted by Schrodinger \cite{schrod} in that it allowed
for the steering of the state on one side merely  by the choice
of the measurement basis on the other side without in any way having direct
access to the affected particle. The word `entanglement' was first coined
by Schrodinger to describe the property of such spatially separated but
correlated particles.  

Inspired by the early works of EPR and Schrodinger, a formalism for 
quantifying the correlations in terms of joint measurements of observables  
corresponding to two spatially separated particles was first proposed by
Bell \cite{bell} for the case of any general theory obeying the tenets of 
locality and realism. Bell's inequality was shown to be violated in
quantum mechanics, a fact that has since been empirically validated in several
subsequent experiments \cite{aspect}. From the practical point of view, 
quantum correlations have been used as resource in 
performing tasks that are unable to be achieved using classical means,
leading to many interesting and important information theoretic applications, 
such as dense coding, teleportation, and secret key generation.
Developments in quantum information theory for both discrete \cite{QI_Disc_09} as well as continuous
variables \cite{QI_Conti_05}  have brought about the realization
of subtle differences in various categories of correlations, for example,
the distinction of quantum entanglement from more general quantum correlations,
{\it viz.}, quantum discord \cite{discord} found in classes of 
separable states. 

On the other hand, the understanding of the precise nature of correlations
that lead to the EPR paradox had to wait for a number of years beyond 
Bell's derivation of his inequality, and further advances in quantum 
information theory. In this direction, a testable formulation of the EPR 
paradox was proposed by Reid \cite{reid} in the realm of continuous variable
systems using the position-momentum uncertainty relation, in terms of an 
inequality involving products of inferred variances of incompatible 
observables. This lead to the
experimental realization of the EPR paradox by Ou et al. \cite{ou} for
the case of two spatially separated and correlated light modes. Similar
demonstrations of the EPR paradox using quadrature amplitudes of other 
radiation fields were performed \cite{tara}.  Moreover, a
much stronger violation of the Reid inequality for two mode squeezed vaccum
states has been experimentally demonstrated recently \cite{stein}. 
The EPR criterion has been used to demonstrate entanglement in Bose-Einstein
condensates, as well \cite{he}.
Other works have showed that
the Reid inequality is effective in demonstrating the EPR paradox for
systems in which correlations appear at the level of variances. 
However, in systems
with correlations manifesting in higher than the second moment, the Reid
formulation generally fails to show occurrence  of the EPR paradox, even though
Bell nonlocality may be exhibited \cite{walborn,lg}.

A more direct manifestation of EPR-type correlations has been proposed by the 
work of Wiseman et al. \cite{wiseman1,wiseman2}, where steering is formulated 
in terms
of an information theoretic task. Using similar formulations for entanglement
as well as Bell nonlocality, a clear distinction between these three types
of correlations is possible using joint probability distributions. Wiseman
et al. \cite{wiseman1,wiseman2} have further shown a hierarchy
between the three types of correlations, with entanglement being the
weakest, steering the intermediate, and Bell violation the strongest
of the three. Bell nonlocal states constitute a strict subset of steerable
states which, in turn, are a strict subset of entangled states. For the case 
of pure entangled
states of two qubits the three classes overlap. An experimental
demonstration of these differences has been demonstrated for mixed
entangled states of two qubits \cite{saunders}. A loophole free EPR-steering
experiment has also been performed \cite{witt}. 
The case of continuous
variable states however poses an additional difficulty, since there exist
several pure entangled states which do not display steering through the
Reid criterion based on variances of observables \cite{reid}. In order to 
exploit higher order correlation
in such states, Walborn et al. \cite{walborn} proposed a new steering
condition which is derived using the entropic uncertainty principle 
\cite{bialynicki}. Entropic functions by definition incorporate correlations 
up to all orders, and  the Reid criterion can be seen to follow as a
limiting case of the entropic steering relation \cite{walborn}. Generalizations
of entropic steering inequalities to the case of symmetric 
steering \cite{schn}, loss-tolerant steering \cite{Loss_toler}, as
well as to the case of steering with qauntum memories \cite{schn2} has
also been proposed recently.     

EPR steering for Gaussian states has been studied extensively both
theoretically and experimentally. It is realized though that Gaussian states
are a rather special class of states, and there exist very common examples
of states, such as the superposition of two oscillators in Fock states that
are far from Gaussian in nature. The non-Gaussian states are usually generated 
by the process of photon subtraction and addition \cite{book}, and these 
states generally have higher degree of entanglement than the Gaussian states. 
Hence, non-Gaussian states have applications in tests of Bell inequalities, 
quantum teleportation and other quantum information protocols \cite{qinfo}. 
Extensions of the entanglement criteria for non-Gaussian
states have been proposed recently \cite{ENT_NG}.
Since the steering of correlated systems has started being studied only 
recently, it is important to understand the steering of systems with 
non-Gaussian correlations. A particular example of a non-Gaussian state
was considered by Walborn et al. \cite{walborn} revealing steering through
the entropic inequality.  
Non-Gaussian entanglement and steering has also been
recently studied in the context of Kerr-squeezed optical
beams \cite{Olsen_2013}.
In the present work we consider several categories of non-Gaussian states
with the motivation to investigate EPR-steering of such states. This
should stimulate steering experiments using non-Gaussian
states.

The plan of the paper is as follows. In the next section we present a 
brief review of the basic concepts involved in EPR steering. Here we
first discuss the Reid criterion for demonstrating the EPR paradox,
and recall its applicability for the case of the two mode squeezed vacuum
state. We then discuss steering as an information processing task,
and the entropic steering inequality for conjugate variable pairs.
In Section III, several examples of non-Gaussian states are studied
for their steering and nonlocality properties. Here we first consider
entangled eigenstates of the two dimensional harmonic oscillator given
by Laguerre-Gaussian wave functions that have been experimentally 
realized \cite{review,fickler}, and may be capable  of useful information 
processing due to their
high available degrees of freedom. We show the inadequacy of
the Reid criterion in revealing steering for such states. We
then demonstrate steering using the entropic steering relation. 
Photon subtraction from light beams is useful for generating a variety
of non-Gaussian states, and 
is thought to be of much practical use in quantum state engineering 
\cite{zavatta}. We next study the steering properties of photon
subtracted squeezed vacuum states using the entropic steering inequality. 
Lastly, we study steering by N00N states \cite{dow} that are regarded to
be of high utility in quantum metrology. In all the examples considered,
we present a comparison of the magnitude of Bell violation with the
strength of steering. Such an analysis also brings out the
comparative efficiency of the  steering framework in revealing quantum
correlations in a given state compared to the Bell framework,
that may be of practical relevance. A summary of our main results are
presented in Section IV.

\section{The EPR paradox and steering}

The EPR paradox may be understood by considering a bipartite entangled
state which may be expressed in two different ways, as
\begin{eqnarray}
\vert\Psi\rangle = \sum_{n=1}^{\infty}c_n\vert\psi_n\rangle\vert u_n\rangle =
 \sum_{n=1}^{\infty}d_n\vert\phi_n\rangle\vert v_n\rangle
\label{ensemb}
\end{eqnarray}
where $\{\vert u_n\rangle\}$ and $\{\vert v_n\rangle\}$ are two orthonormal
bases for one of the parties (say, Alice). If Alice chose to measure in the 
$\{\vert u_n\rangle\}$  ($\{\vert v_n\rangle\}$) basis, then she 
instantaneously projects
Bob's system into one of the states $\vert\psi_n\rangle$ ($\vert\phi_n\rangle$).
This ability of Alice to affect Bob's state due to her choice of the 
measurement basis was dubbed as ``steering'' by Schrodinger \cite{schrod}.
Since there is no physical interaction between Alice and Bob, it is paradoxical
that the ensemble of $\vert\psi_n\rangle$s is different from the ensemble
of $\vert\phi_n\rangle$s.

The
EPR paradox stems from the correlations between two non-commuting
observables of a sub-system with those of the other sub-system, i.e.,
$<x,p_y> \neq 0$, with $<x>=0=<p_y>$ individually. 
In the original formulation of the paradox correlations between the measurement outcomes
of positions and momenta for two separated particles was considered. Due to the presence of correlations,
the measurement of the position of, say, the first particle leads one to infer the correlated value of
the position for the second particle (say, $x_{\inf}$). Now, if the momentum of the second particle is measured
giving the outcome, say $p$, the value of the product of uncertainties $(\Delta x_{\inf})^2 (\Delta p_{\inf})^2$ may
 turn out to be lesser than that allowed by the uncertainty principle, {\it viz.} $(\Delta x)^2 (\Delta p )^2 \ge 1$, thus leading to the paradox.
The following material in this section is primarily to fix the setting for
our work on non-Gaussian entangled states.

\subsection{The Reid inequality and its violation for the two mode squeezed vacuum state}

The possibility of demonstrating the EPR paradox in the context
 of continuous variable correlations was first proposed by Reid \cite{reid}. Such an idea has been experimentally realized \cite{ou} through quadrature phase measurements performed on the two output
 beams of a nondegenerate parametric amplifier. This technique of demonstrating the product of variances of the inferred values of
 correlated observables to be less than that allowed by the uncertainty principle, has since gained popularity \cite{tara}, and has been employed recently for variables other than position and momentum, e.g., for correlations between
 optical and orbital angular momentum of light emitted through
 spontaneous parametric down-conversion \cite{leach}.

Let us now consider the situation where the quadrature phase components of two correlated and spatially
separated light fields are measured.  The quadrature amplitudes associated with the fields $E_{\gamma}=C[\hat{\gamma} e^{-i\omega_{\gamma} t} + \hat{\gamma}^{\dagger} e^{i\omega_{\gamma} t}]$ (where, $\gamma\in\{a,b\}$, are the bosonic operators for two different modes, $\omega_{\gamma}$ is the frequency, and
$C$ is a constant incorporating spatial factors taken to be equal for each mode) are given by
\begin{eqnarray}
\hat{X}_{\theta}=\frac{\hat{a}e^{- i \theta} + \hat{a}^{\dagger} e^{i \theta}}{\sqrt{2}},
\hspace{0.5cm}
\hat{Y}_{\phi}=\frac{\hat{b}e^{- i \phi} + \hat{b}^{\dagger} e^{i \phi}}{\sqrt{2}},
\label{Quard}
\end{eqnarray}
where,
\begin{eqnarray}
\hat{a} &=& \frac{X + i P_x}{\sqrt{2}},\hspace{0.5cm} \hat{a}^\dagger = \frac{X -i P_x}{\sqrt{2}},\nonumber\\
\hat{b}&=&  \frac{Y+i P_y}{\sqrt{2}}, \hspace{0.5cm} \hat{b}^\dagger = \frac{Y- i P_y}{\sqrt{2}},
\label{boson_op}
\end{eqnarray}
and the commutation relations of the bosonic operators are given by $[\hat{a},\hat{a}^{\dagger}]=1=[\hat{b},\hat{b}^{\dagger}]$.
Now, using Eq.(\ref{boson_op})  the expression for the quadratures can be rewritten as
\begin{eqnarray}
\hat{X}_{\theta} = \cos[\theta] ~\hat{X} + \sin[\theta] ~\hat{P}_x, \hspace{0.3cm}
\hat{Y}_{\phi} = \cos[\phi]~ \hat{Y} +\sin[\phi]~ \hat{P}_y.
\label{Dless}
\end{eqnarray}
The correlations between the quadrature amplitudes $\hat{X}_{\theta}$ and $\hat{Y}_{\phi}$ are captured by the correlation coefficient, $ C_{\theta,\phi} $  defined as \cite{reid,ou,tara}
\begin{eqnarray}
C_{\theta,\phi}=\frac{\langle \hat{X}_{\theta} \hat{Y}_{\phi} \rangle}{\sqrt{\langle \hat{X}^2_{\theta} \rangle \langle \hat{Y}^2_{\phi}  \rangle}},
\label{Cr_f}
\end{eqnarray}
where $\langle \hat{X}_{\theta} \rangle=0=\langle \hat{Y}_{\phi} \rangle$. The correlation is perfect for some values of $\theta$ and $\phi$, if $|C_{\theta,\phi}|=1$. Clearly $|C_{\theta,\phi}|=0$ for uncorrelated variables.

Due to the presence of correlations, the quadrature amplitude $\hat{X}_{\theta}$ can be inferred by measuring the corresponding amplitude $\hat{Y}_{\phi}$. The
 EPR paradox arises due to the ability to infer an observable of one system from the result of measurement
 performed on a spatially separated second system. In realistic situations the correlations are not perfect because of the interaction with the environment as well as finite detector efficiency. Hence, the estimated amplitudes $\hat{X}_{\theta 1}$ and $\hat{X}_{\theta 2}$ with the help of $\hat{Y}_{\phi 1}$ and $\hat{Y}_{\phi 2}$, respectively, are subject to inference errors, and given by \cite{reid}
\begin{eqnarray}
\hat{X}_{\theta 1}^{e}=g_1 \hat{Y}_{\phi 1},
\hspace{0.5cm}
\hat{X}_{\theta 2}^{e}=g_2 \hat{Y}_{\phi 2},
\label{Est}
\end{eqnarray}
where $g_1$ and $g_2$ are scaling parameters.
Now, one may choose $g_1$, $g_2$, $\phi 1$, and $\phi 2$ in such a way that $\hat{X}_{\theta 1}$ and $\hat{X}_{\theta 2}$ are inferred with the highest possible accuracy. The errors given by the deviation of the estimated amplitudes from the true amplitudes $\hat{X}_{\theta 1}$ and $\hat{X}_{\theta 2}$ are captured by $(\hat{X}_{\theta 1}- \hat{X}_{\theta 1}^{e})$ and $(\hat{X}_{\theta 2}- \hat{X}_{\theta 2}^{e})$, respectively. The average errors of the inferences are given by
\begin{eqnarray}
(\Delta_{\inf} \hat{X}_{\theta 1})^2 &=& \langle (\hat{X}_{\theta 1}- \hat{X}_{\theta 1}^{e})^2\rangle =  \langle (\hat{X}_{\theta 1}- g_1 \hat{Y}_{\phi 1})^2\rangle, \nonumber \\
(\Delta_{\inf} \hat{X}_{\theta 2})^2 &=& \langle (\hat{X}_{\theta 2}- \hat{X}_{\theta 2}^{e})^2\rangle =  \langle (\hat{X}_{\theta 2}- g_2 \hat{Y}_{\phi 2})^2\rangle.
\label{Erro}
\end{eqnarray}
The values of the scaling parameters $g_1$ and $g_2$ are chosen such that
$\frac{\partial (\Delta_{\inf} \hat{X}_{\theta 1})^2}{\partial g_1} =0 = \frac{\partial (\Delta_{\inf} \hat{X}_{\theta 2})^2}{\partial g_2}$, from which it follows that
\begin{eqnarray}
g_1 = \frac{\langle \hat{X}_{\theta 1} \hat{Y}_{\phi 1} \rangle}{\langle \hat{Y}_{\phi 1}^2 \rangle},
\hspace{0.5cm}
g_2 = \frac{\langle \hat{X}_{\theta 2} \hat{Y}_{\phi 2} \rangle}{\langle \hat{Y}_{\phi 2}^2 \rangle}.
\label{g's}
\end{eqnarray}
The values of $\phi 1$ ($\phi2$) are obtained by maximizing $C_{\theta1,\phi1}$ ($C_{\theta 2,\phi2}$).
 Now, due to the commutation relations $[\hat{X},\hat{P}_X]=i;~~[\hat{Y},\hat{P}_Y]=i$, it is required
 that the product
 of the variances of the above inferences $(\Delta_{\inf} \hat{X}_{\theta 1})^2 (\Delta_{\inf} \hat{X}_{\theta 2})^2 \ge 1/4$. Hence, the EPR paradox occurs if the correlations in the field quadratures lead to
 the condition
 \begin{eqnarray}
EPR \equiv (\Delta_{\inf} \hat{X}_{\theta 1})^2 (\Delta_{\inf} \hat{X}_{\theta 2})^2 < \frac{1}{4}.
\label{P_Uncer}
\end{eqnarray}

Let us consider a two mode squeezed vacuum (TMSV) state, the expression of which is given by\cite{NOPA}
\begin{eqnarray}
|NOPA\rangle = |\xi\rangle &=& S(\xi) |0,0\rangle  \nonumber\\
& = & \sqrt{1-\lambda^2}~~\displaystyle\sum_{n=0}^{\infty}\lambda^n~|n,n\rangle
\label{NOPA}
\end{eqnarray}
where, $ \lambda=\tanh(r)\in[0,1] $, the squeezing parameter $ r>0 $ and $ |m,n\rangle=|m\rangle_A\otimes |n\rangle_B $ (where $|m\rangle$ and $|n\rangle$ are the usual Fock states). $S(\xi)$ ($=e^{(\xi\hat{a_1}^{\dagger}\hat{a_2} ^{\dagger}-\xi^\ast a_1 a_2)}$, where $\xi=r e^{i\phi}$) is the squeezing operator (unitary). $ A $ and $ B $ are the two involved modes for Alice and Bob respectively.

For the NOPA state given by Eq.(\ref{NOPA}), the inferred uncertainties is given by
\begin{eqnarray}
(\Delta_{\inf} X_{\theta})^2&=&\frac{1}{2} \cosh[2r] \nonumber \\ 
&& - \frac{1}{2} \tanh[2r] \sinh[2r] \cos^2[\theta+\phi],
\end{eqnarray}
where the quadrature amplitude $X_\theta$ is inferred by measuring the corresponding amplitude $Y_\phi$.
The minimum values for two different values of $\theta $ (i. e. , $\theta_1=0$ and $\theta_2=\pi/2$) of  $(\Delta_{\inf} X_{\theta})^2$ are
\begin{eqnarray}
(\Delta_{\inf} X_{\theta_1})^2= (\Delta_{\inf} X_{\theta_2})^2= \frac{1}{2 \cosh[2r]} ,
\label{NOPA_Reid}
\end{eqnarray}
which occur for $\phi_1=0$ and $\phi_2=\pi/2$, respectively. The product of
uncertainties  is thus $\frac{1}{4 \cosh^2[2r]}$ which asymptotically reaches the value $0$ for $ r\rightarrow \infty $, and this shows that the Reid
condition (\ref{P_Uncer}) for occurrence of the EPR paradox holds. Hence, 
the two mode squeezed vacuum state shows EPR steering for all values of $ r $ except at $ r=0$. However, the Reid condition fails to demonstrate
steering by more general non-Gaussian states, for example, the two-dimensional
harmonic oscillator, as we will show in Section III.

\subsection{Steering and entropic inequalities}

A modern formulation of EPR steering was presented by Wiseman et 
al. \cite{wiseman1,wiseman2} as an information processing task. They considered
that one of two parties (say, Alice) prepares a bipartite quantum state
and sends one of the particles to Bob. The procedure is repeated as many
times as required. Bob's particle is assumed to possess a definite
state, even if it is unknown to him (local hidden state). No such 
assumption is made for Alice, and hence, this
formulation of steering is an asymmetric task. 
Alice and Bob make measurements on their respective 
particles, and communicate classically. Alice's task is to convince Bob
that the state they share is entangled. If correlations between Bob's 
measurement results and Alice's  declared results can be explained by 
a local hidden state (LHS) model for Bob, he is not convinced. This is
because Alice could have drawn a pure state at random from some ensemble and 
sent it to Bob, and then chosen her result based on her knowledge of this LHS.
 Conversely, if the correlations cannot be so explained, then the state must 
be entangled. Alice will be successful in her task of steering if she can 
create genuinely different ensembles for Bob by steering Bob's state. It may
be noted that a similar formulation of Bell nonlocality as an information 
theoretic task is also possible \cite{wiseman1}, where the correlations
between Alice and Bob may be described in terms of a local hidden variable
model. 

In the above situation, an EPR-steering inequality \cite{caval} may be
derived involving an experimental situation for qubits with $n$ 
measurement settings
for each side. Bob's $k$-th measurement
setting is taken to correspond with the observable $\hat{\sigma}_k$, 
and Alice's declared result is denoted by the random variable 
$A_k \to \{-1,1\}$. Violation of the inequality
\begin{eqnarray}
\frac{1}{n}\sum_{k=1}^n\langle A_k\hat{\sigma}_k\rangle \le C_n
\label{steer1}
\end{eqnarray}
reveals occurrence of steering, where $C_n \equiv \mathrm{max}_{\{A_k\}}(\frac{\lambda_{\mathrm{max}}}{n}\sum_{k=1}^nA_k\hat{\sigma}_k)$ is the maximum
value of the l.h.s. of (\ref{steer1}) if Bob has a pre-existing state known 
to Alice, with $\lambda_{\mathrm{max}}$ being the largest eigenvalue of
the operator $\frac{1}{n}\sum_{k=1}^n A_k\hat{\sigma}_k$. Experimental 
demonstration of steering for mixed entangled states \cite{saunders}
that are Bell local has confirmed that steering is a weaker form of
correlations compared to nonlocality.

For the case of continuous variable systems, the Reid criterion is an
indicator for steering, as discussed above. However, there exist several
pure entangled continuous variable states which do not reveal steering
through the Reid criterion. An example of such a state is provided in
Ref.\cite{walborn}, which we also discuss briefly below. Since entanglement 
is a weaker form of correlations
compared to steering \cite{wiseman1,wiseman2}, it is clear that for such
states the steering correlations do not appear up to second order (variances)
that may be checked by the Reid criterion. The Reid criterion itself is
derived using the Heisenberg uncertainty relation involving product of
variances of non-commuting observables. On the other hand, a more general 
form of the uncertainty relation containing correlations in all orders
of, for example, the position and momentum distribution of a quantum system is
provided by the entropic uncertainty relation \cite{bialynicki} given by
\begin{eqnarray}
h_Q(X)+h_Q(P)\geq \ln \pi e.
\label{entropy_uncertainty}
\end{eqnarray}
Using the entropic uncertainty relation, Walborn et al. \cite{walborn} have
derived an entropic steering inequality. They considered a joint probability
distribution of two parties corresponding to a non-steerable state for
which there exists a local hidden state (LHS) description, given by
\begin{eqnarray}
\mathcal{P}(r_A,r_B)=\sum_\lambda \mathcal{P}(\lambda)\mathcal{P}(r_A|\lambda)\mathcal{P}_Q(r_B|\lambda),
\label{steer2}
\end{eqnarray}
where, $ r_A $ and $ r_B $ are the outcomes of measurements $ R_A $ and $ R_B $ respectively;   $ \lambda $ are hidden variables that specify an ensemble of 
states; $ \mathcal{P} $ are general probability distributions; and $ \mathcal{P}_Q $ are probability distributions corresponding to the quantum state specified by $ \lambda $. Now, using a rule for conditional probabilities
$P(a,b|c) = P(b|c)P(a|b)$ which holds when $\{b\} \in \{c\}$, i.e., there 
exists a local hidden state of Bob predetermined by Alice, it follows that
the conditional probability $\mathcal{P}(r_B| r_A)$ is given by
\begin{eqnarray}
\mathcal{P}(r_B|r_A)=\sum_\lambda \mathcal{P}(r_B,\lambda|r_A)
\label{steer3}
\end{eqnarray}
with $P(r_B,\lambda | r_A) = P(\lambda |r_A)P_Q(r_B|\lambda)$. Note that
(\ref{steer2}) and (\ref{steer3}) are equivalent conditions for 
non-steerability. Next, considering the relative entropy (defined for two
distributions $p(X)$ and $q(X)$ as $\mathcal{H}(p(X)||q(X))= \sum_xp_x\ln(p_x/q_x)$) between the
probability distributions $ \mathcal{P}(r_B,\lambda|r_A) $ and $ \mathcal{P}(\lambda|r_A)\mathcal{P}(r_B|r_A) $ , it follows from the positivity of
relative entropy that
\begin{eqnarray}
\sum_\lambda \int dr_B \mathcal{P}(r_B,\lambda|r_A) \ln \frac{\mathcal{P}(r_B,\lambda|r_A)}{\mathcal{P}(\lambda|r_A)\mathcal{P}(r_B|r_A)}\geq 0
\end{eqnarray}
Using the non-steering condition (\ref{steer3}), the definition of the 
conditional entropy ($h(X|Y) = -\sum_{x,y} p(x,y)\ln p(x|y)$), and averaging over all measurement 
outcomes $r_A$, it
follows that the conditional entropy $h(R_B|R_A)$ satisfies
\begin{eqnarray}
h(R_B|R_A) \ge \sum_{\lambda} \mathcal{P}(\lambda) h_Q(R_B|\lambda)
\label{cond1}
\end{eqnarray}
Considering a pair of variables $S_A,S_B$ conjugate to $R_A,R_B$, a similar
bound on the conditional entropy may be written as
\begin{eqnarray}
h(S_B|S_A) \ge \sum_{\lambda} \mathcal{P}(\lambda) h_Q(S_B|\lambda)
\label{cond2}
\end{eqnarray}
For the LHS model for Bob, note that the entropic uncertainty relation 
(\ref{entropy_uncertainty}) holds for each state marked by $\lambda$. 
Averaging over all hidden variables, it follows that
\begin{eqnarray}
 \sum_{\lambda} \mathcal{P}(\lambda)\biggl(h_Q(R_B|\lambda)
+ h_Q(S_B|\lambda)\biggr) \ge \ln \pi e
\label{cond3}
\end{eqnarray} 
Now, using the bounds (\ref{cond1}) and (\ref{cond2}) in the relation
(\ref{cond3}) one gets the entropic steering inequality given by
\begin{eqnarray}
h(R_B|R_A)+h(S_B|S_A)\geq \ln \pi e.
\label{entropy_steering}
\end{eqnarray}
Walborn et al. \cite{walborn} presented an example of the state given by
(up to a suitable normalization)
\begin{eqnarray}
\phi_n(x_A,x_B) = \mathcal{H}_n(\frac{x_A+x_B}{\sqrt{2}\sigma_{+}})e^{-\frac{(x_A+x_B)^2}{4\sigma_{+}^2}}e^{-\frac{(x_A-x_B)^2}{4\sigma_{-}^2}}
\label{walex}
\end{eqnarray}
where $\mathcal{H}_n$ is the $n$-th order Hermite polynomial, which does not
reveal steering using the Reid criterion when $\sigma_{\pm}/\sigma_{\mp} < 1 + 1.5\sqrt{n}$, whereas the entropic steering criterion (\ref{entropy_steering})
is able to show steering except when the state is separable, i.e., for
$n=0$, and $\sigma_{+} = \sigma_{-}$.
Using the relation between information entropy and variance, it was further
shown by Walborn et al. \cite{walborn} that the Reid criterion follows
in the limiting case of the entropic steering relation (\ref{entropy_steering}).
In the following section we will use the entropic steering inequality
for demonstrating steering by several continuous variable states.

\section{Steering and nonlocality for non-Gaussian states}

In this section we study steering and nonlocality by several non-Gaussian
states. Considering first entangled states constructed using the
eigenstates of the two-dimensional harmonic oscillator, we study the steering
and nonlocal properties of LG beams. We show that the Reid criterion is
unable to reveal the streerability of LG modes. The entropic steering 
inequality shows that the strength of steering increases with angular
momentum of the LG beams. We then discuss non-Gaussian states
obtained by subtracting single and two photons from two-mode squeezed
vacuum states. We show that the violation of Bell's inequality for
such states behaves differently with the increase of the squeezing parameter
compared to the strength of steering. Finally, we investigate the nonlocal
and steering properties of another class of non-Gaussian states, {\it viz.},
N00N states.

\subsection{Non-Gaussian entangled states of a two dimensional harmonic oscillator}

The importance of the two-dimensional harmonic oscillator cannot be
overemphasized in the context of quantum mechanics. The historical development
of radiation theory started with the correspondence with the two modes of
the radiation field. The classic problem of the charged particle in
the electromagnetic field leading to the existence of Landau levels
was developed using the same machinery. The energy eigenfunctions of
the two-dimensional harmonic oscillator may be expressed in terms of
Hermite-Gaussian (HG) functions given by 
\begin{eqnarray}
u_{nm}(x,y) &&= \sqrt{\frac{2}{\pi}} \left(\frac{1}{2^{n+m} w^2 n!m!}\right)^{1/2} \nonumber\\
&& \times H_n \left(\frac{\sqrt{2}x}{w}\right) H_m \left(\frac{\sqrt{2}y}{w}\right) e^{-\frac{(x^2+y^2)}{w^2}},
\nonumber \\
 \int |u_{nm}(x,y)|^2 dx dy &&=1
\label{hermite}
\end{eqnarray}
Entangled states may be constructed  from superpositions of
 HG wave functions \cite{danakas}
\begin{eqnarray}
\Phi_{nm}(\rho,\theta) = \sum_{k=0}^{n+m} u_{n+m-k,k}(x,y)\frac{f_k^{(n,m)}}{k!}(\sqrt{-1})^k \nonumber \\
\times \sqrt{\frac{k! (n+m-k)!}{n! m! 2^{n+m}}}
\label{legherm}
\end{eqnarray}
\begin{eqnarray}
f_k^{(n,m)} = \frac{d^k}{dt^k} ((1-t)^n(1+t)^m)|_{t=0},
\label{def11}
\end{eqnarray}
where $\Phi_{nm}(\rho,\theta)$ are the well-known Laguerre-Gaussian functions
that are physically realizable field configurations \cite{review,fickler} 
with interesting topological \cite{berry} and coherence \cite{simon,banerji}
properties, given by
\cite{book2}
\begin{eqnarray}
\Phi_{nm}(\rho,\theta) = e^{i(n-m)\theta}e^{-\rho^2/w^2}(-1)^{\mathrm{min}(n,m)}
\left(\frac{\rho \sqrt{2}}{w}\right)^{|n-m|} 
\label{waveLG} \\
\times \sqrt{\frac{2}{\pi n! m ! w^2}}
L^{|n-m|}_{\mathrm{min}(n,m)} \left(\frac{2\rho^2}{w^2}\right) (\mathrm{min}(n,m)) ! \nonumber
\end{eqnarray}
with
 $\int |\Phi_{nm}(\rho,\theta)|^2 dx dy =1$,
where $w$ is the beam waist, and $L_p^l(x)$ is the generalized Laguerre polynomial. 
The superposition (\ref{legherm}) is like a Schmidt decomposition
thereby
signifying the entanglement of the LG wave functions.
In the special case
\begin{eqnarray}
\Phi_{10} = \frac{2}{\sqrt{\pi} w^2} (x + iy)e^{-(x^2+y^2)/w^2} \nonumber\\
\Phi_{01} = \frac{2}{\sqrt{\pi} w^2} (x - iy)e^{-(x^2+y^2)/w^2}
\label{zerolg} 
\end{eqnarray}
In the following analysis we will study the quantum correlations present
in the LG wave functions, first for the purpose of demonstrating steering.
We will next compare the strength of steering with the degree of nonlocality
of such modes through the violation of Bell's inequality \cite{bell}.

It is henceforth convenient to work with the pair of dimensionless quadratures
$\{X,~P_X\}$ and $\{Y,~P_Y\}$, given by
\begin{eqnarray}
x (y) \rightarrow \frac{w}{\sqrt{2}} ~~X (Y),
\hspace{0.5cm}
p_x (p_y) \rightarrow \frac{\sqrt{2} \hbar}{w} ~~P_X (P_Y),
\label{DL_T}
\end{eqnarray}
The canonical commutation relations are $[\hat{X},\hat{P}_X]=i;~~[\hat{Y},\hat{P}_Y]=i$, and the operator $\hat{P}_X$ and $\hat{P}_Y$ are given by
$\hat{P}_X = - i \frac{\partial}{\partial X}$ and $\hat{P}_Y = - i \frac{\partial}{\partial Y}$, respectively. The Wigner function corresponding to the LG wave function
  in terms of the scaled variables is given by
\begin{eqnarray}
W_{nm}(X,P_X;Y,P_Y)&=&\frac{(-1)^{n+m}}{(\pi)^{2}} L_{n}[4(Q_0+Q_2)]
\label{WF_LG_n1_n2} \\
&& L_{m}[4(Q_0-Q_2)]~exp(-4Q_0) \nonumber
\end{eqnarray}
where
\begin{eqnarray}
Q_0 & = & \frac{1}{4}\left[ X^2 + Y^2 + P_X^2+P_Y^2\right],\\
Q_2 & = &  \frac{XP_Y-YP_X}{2}.
\label{wig}
\end{eqnarray}

\begin{figure}[!ht]
\resizebox{9cm}{6cm}{\includegraphics{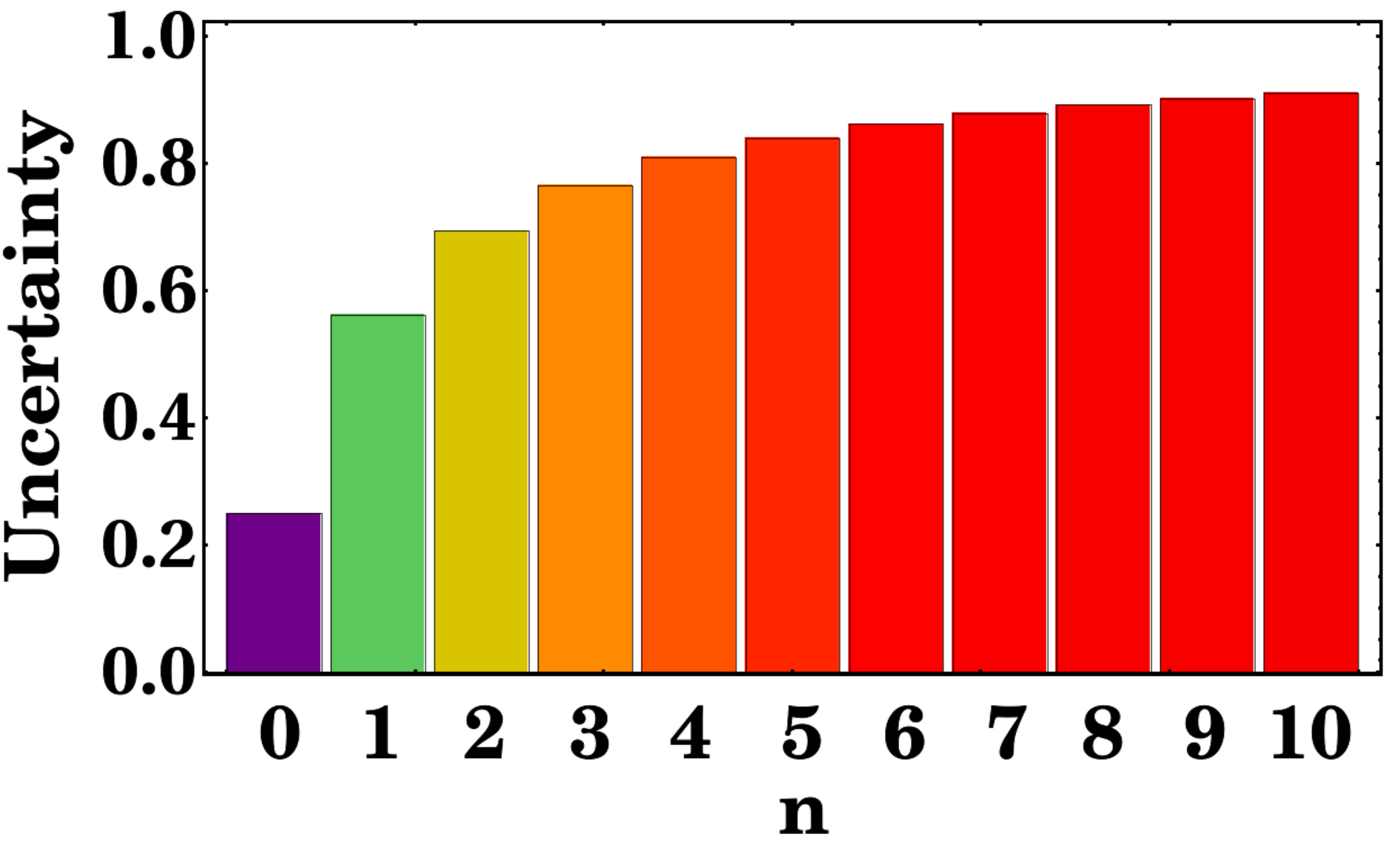}}
\caption{\footnotesize (Coloronline)
The product of uncertainties $(\Delta_{\inf} X_{\theta_1})^2(\Delta_{\inf} X_{\theta_2})^2$ is plotted versus $n$ for $m=0$. The figure shows that LG beam does not demonstrate steering through the Reid
criterion.
}
\label{LG_Steer_Reid}
\end{figure}

Let us now check how the Reid criterion applies to the case of LG wave fucntions.
In order to do so we estimate the product of uncertainties of the values
of inferred observables $(\Delta_{\inf} X_{\theta_1})^2(\Delta_{\inf} X_{\theta_2})^2$.
This is performed by maximizing the correlation function 
$C_{\theta1,\phi1}$ ($C_{\theta 2,\phi2}$). Using Eqs.(\ref{Erro}) and (\ref{g's})
it follows that \begin{eqnarray}
(\Delta_{\inf} X_{\theta})^2 = \langle X_\theta^2 \rangle \left(1 - (C^{\max}_{\theta,\phi})^2\right)
\label{X_Unc}
\end{eqnarray}
The maximum correlation strength $|C^{\max}_{\theta,\phi}| ~=\frac{1}{2}$ occurs for $\phi-\theta=\frac{k \pi}{2} $ (where $k$ is an odd integer). For arbitrary values of $n,m$ it can be shown that the
expression of the maximum correlation function is given by
\begin{eqnarray}
C^{\max}_{0,\pi/2} = \frac{\langle XP_Y \rangle}{\sqrt{\langle X^2 \rangle \langle P^2_Y\rangle}},~~~ C^{\max}_{\pi/2,\pi} =
- \frac{\langle P_X Y \rangle}{\sqrt{\langle P_X^2 \rangle \langle Y^2\rangle}}
\label{maxcor}
\end{eqnarray}
In Figure-\ref{LG_Steer_Reid} we plot the product of uncertainties 
$(\Delta_{\inf} X_{\theta_1})^2(\Delta_{\inf} X_{\theta_2})^2$ versus the 
angular momentum $n$. It is seen that the Reid criterion given by 
Eq.(\ref{P_Uncer}) is not satisfied, since 
$(\Delta_{\inf} X_{\theta_1})^2(\Delta_{\inf} X_{\theta_2})^2 \ge 1/4$  for any value of $n$. Hence, it
is not possible to demonstrate steering by entangled LG modes using the
Reid criterion.

We now apply the entropic steering criterion to the case of the LG wave functions.
In the entropic steering inequality given by Eq.(\ref{entropy_steering})
the observables have to be chosen such that there exist correlations between
$R_A$ and $R_B$ ($S_A$ and $S_B$). For the case of the LG wave functions, we use
the nonvanishing $\langle X P_Y\rangle$ correlations, as evident from
the Wigner function (\ref{wig}). Thus, in terms of the conjugate pairs of dimensionless quadratures, (\ref{entropy_steering}) becomes
\begin{eqnarray}
h(\mathcal{X}|\mathcal{P_Y})+h(\mathcal{P_X|}\mathcal{Y})\geq \ln \pi e,
\label{LG_entropy_steering}
\end{eqnarray}
where $ X,~Y,~P_X $ and $ P_Y $ are the outcomes of measurements $ \mathcal{X},~\mathcal{Y},~\mathcal{P_X} $ and $ \mathcal{P_Y} $ respectively.
Here, the conditional entropies $ h(\mathcal{X}|\mathcal{P_Y}) $ and $ h(\mathcal{P_X|}\mathcal{Y}) $ are given by
\begin{eqnarray}
h(\mathcal{X}|\mathcal{P_Y})&=& h(\mathcal{X},\mathcal{P_Y})-h(\mathcal{P_Y}),\nonumber\\
h(\mathcal{P_X|}\mathcal{Y})&=&h(\mathcal{P_X},\mathcal{Y})-h(\mathcal{Y}),
\end{eqnarray}
with
$h(\mathcal{X},\mathcal{P_Y})=-\int_{-\infty}^{\infty} \mathcal{P}(X,P_Y) \ln \mathcal{P}(X,P_Y)~ dX dP_Y$, 
$h(\mathcal{P_Y})=-\int_{-\infty}^{\infty} \mathcal{P}(P_Y) \ln \mathcal{P}(P_Y) ~ dP_Y$, and similarly for 
 $ h(\mathcal{P_X},\mathcal{Y}) $ and $ h(\mathcal{Y}) $. The marginal probability
distributions are obtained using the Wigner function (\ref{wig}) for the
LG wave function. 

\begin{figure}[!ht]
\resizebox{9cm}{6cm}{\includegraphics{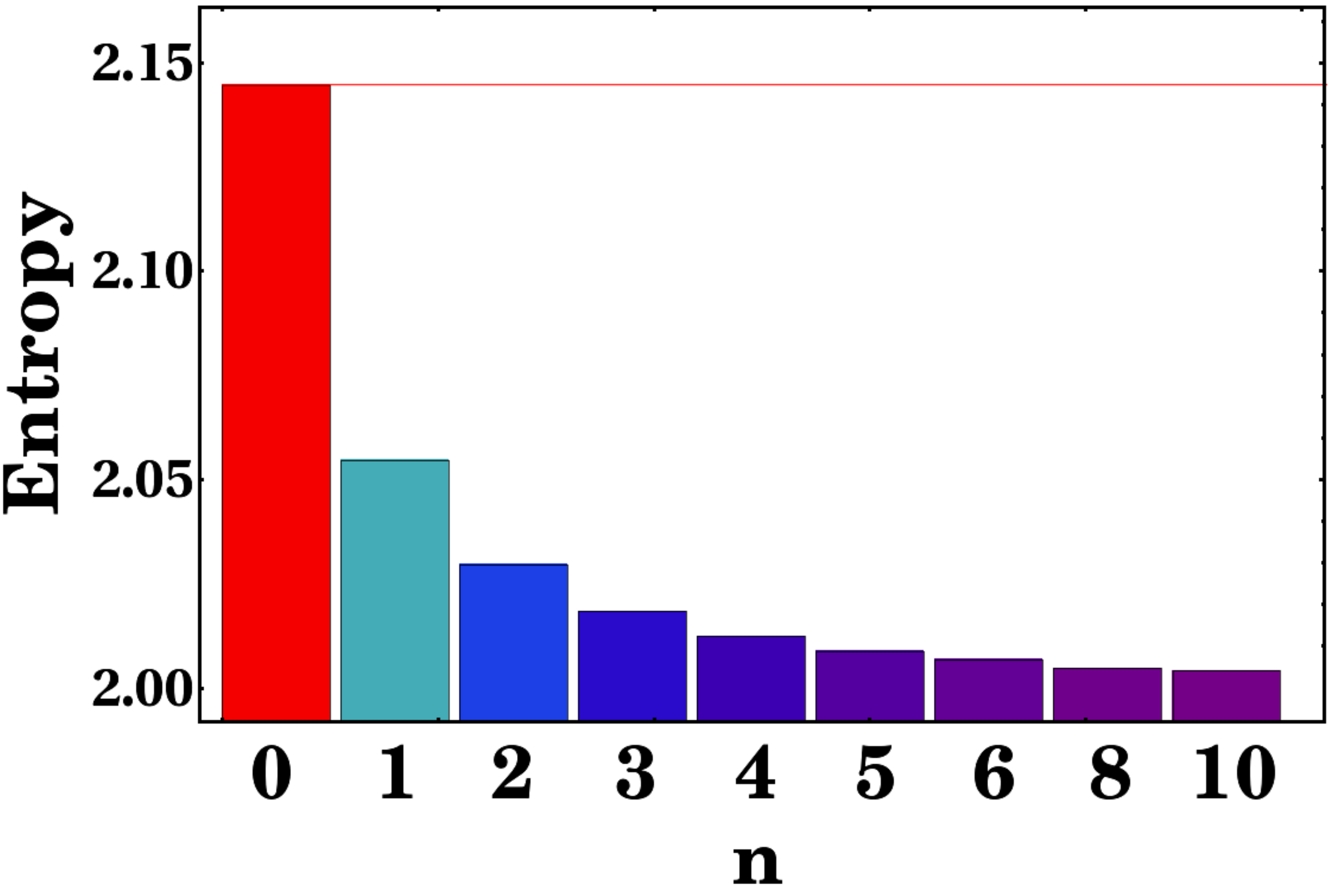}}
\caption{\footnotesize (Coloronline) The figure shows that the violation of entropic steering  inequality (\ref{LG_entropy_steering}) for different values of $n$ (except $n=0$) of the LG wave function keeping $m=0$.  
}
\label{LG_Steer_Fig}
\end{figure}

For $ n=0 $ and $ m=0 $, the LG wave function factorizes into a product
state with the corresponding Wigner function given by
\begin{eqnarray}
W_{00}(X,P_X;Y,P_Y)=\frac{e^{-X^2-Y^2-P_X^2-P_Y^2}}{\pi^2}.
\end{eqnarray}
In this case the relevant entropies turn out to be
$h(\mathcal{X},\mathcal{P_Y})=h(\mathcal{P_X},\mathcal{Y})=\ln \pi e$ and 
$h(\mathcal{Y})=h(\mathcal{P_Y})=\frac{1}{2}\ln \pi e$, and hence, the
entropic steering inequality becomes saturated, i.e.,
\begin{eqnarray}
h(\mathcal{X}|\mathcal{P_Y})+h(\mathcal{P_X|}\mathcal{Y}) = \ln \pi e.
\end{eqnarray}
For $ n=1 $ and $ m=0 $, the Wigner function has the form
\begin{eqnarray}
W_{10}(X,P_X;Y,P_Y)
&=& e^{-X^2-Y^2-P_X^2-P_Y^2}  \\
&& \times \frac{(P_X - Y)^2 +(P_Y+X)^2 -1}{\pi^2} \nonumber
\end{eqnarray}
and the relevant entropies are given by
$h(\mathcal{X},\mathcal{P_Y})=h(\mathcal{P_X},\mathcal{Y}) \approx 2.41509$,
and $h(\mathcal{Y})=h(\mathcal{P_Y}) \approx1.38774$. Hence, the entropic
steering relation in this case becomes
\begin{eqnarray}
h(\mathcal{X}|\mathcal{P_Y})+h(\mathcal{P_X|}\mathcal{Y}) \approx 2.05471 < \ln \pi e 
\end{eqnarray}
We thus see that steering is demonstrated. Note the non-Gaussian nature of
the Wigner function for $n \ge 1$ which enables demonstration of steering
through the entropic criterion. For higher values of angular
momentum, we plot the l.h.s. of the entropic steering relation in 
Figure-(\ref{LG_Steer_Fig}). We see that violation of the inequality
becomes stronger for higher values of $n$.

Now, for making  a comparison between the strength of steering and the
degree of nonlocality, we next study Bell violation by the LG wave function.
In order to do so, we consider the Wigner transform $\Pi_{nm}(X,P_X;Y,P_Y)$ ($=(\pi)^2 \; W_{nm}(X,P_X;Y,P_Y)$, where $W_{nm}(X,P_X;Y,P_Y)$ is given by Eq.(\ref{WF_LG_n1_n2})) \cite{zhang2}. The Bell-CHSH inequality using Wigner transform is given by \cite{paris}
\begin{eqnarray}
|BI| &=& |\Pi_{n,m}(X1,P_{X1};Y1,P_{Y1})\nonumber \\
&& +\Pi_{n,m}(X2,P_{X2};Y1,P_{Y1}) \nonumber \\
&& +\Pi_{n,m}(X1,P_{X1};Y2,P_{Y2})\nonumber \\
&& -\Pi_{n,m}(X2,P_{X2};Y2,P_{Y2})| <2,
\label{BInequality}
\end{eqnarray}
In the following table, we make comparison among Bell violation and entropic EPR steering for different values of $n$  with $m=0$. \\
\begin{tabular}{|c|c|c|c|}
\hline 
  & & & \\
 $n$ & $\frac{|BI_{\max}|}{2}$ & $\frac{(\ln\pi e)}{h(\mathcal{X}|\mathcal{P_Y})+h(\mathcal{P_X|}\mathcal{Y})} $ & $4 ~ (\Delta_{\inf} \hat{X}_{\theta 1})^2 (\Delta_{\inf} \hat{X}_{\theta 2})^2$ \\
   & & & \\
\hline
 0 &    1      & 1  & 1 \\
 \hline
 1 & 1.11934  & 1.04381  & 2.25  \\
 \hline
 2 & 1.17437  &  1.0567 & 2.77778  \\
 \hline
 3 & 1.20128  & 1.06256  & 3.0625  \\
 \hline
 4 & 1.21738  &  1.06572 &  3.24  \\
 \hline
 5 & 1.22813  & 1.06758  & 3.36111  \\
 \hline
 6 & 1.23584  & 1.0687  &  3.44898 \\
 \hline
 7 & 1.24165  & 1.06939  & 3.51563 \\
 \hline
 8 & 1.24618  & 1.0698  & 3.5679  \\
 \hline
 9 & 1.24982 & 1.07002  & 3.61 \\
 \hline
10 & 1.25281  & 1.07011  & 3.64463 \\
 \hline
 \end{tabular} \\

Note here that $\frac{|BI_{\max}|}{2}>1$ signifies Bell violation, and
$\frac{(\ln\pi e)}{h(\mathcal{X}|\mathcal{P_Y})+h(\mathcal{P_X|}\mathcal{Y})} >1$ signifies steering by the entropic steering inequality. On the other hand
the last column provides values of the products of inferred variances,
showing that the Reid criterion is unable to identify steering for any
value of $n$ in this case.

\subsection{Photon subtracted squeezed vacuum}

Let us now consider non-Gaussian states derived from Gaussian states by
the subtraction of photons. Consider the two mode squeezed vacuum state given by
Eq.(\ref{NOPA}). The Wigner function associated with the state (\ref{NOPA}) is given by \cite{book}
\begin{eqnarray}
W_{|\xi\rangle}(\alpha,\beta)&&=\frac{4}{\pi^2} \exp[-2|\alpha \cosh (r) -\beta^{\ast} \sinh(r) \exp[i \phi] |^2 \nonumber \\
&& -2|- \alpha^{\ast} \sinh(r) \exp[i \phi] +\beta \cosh(r) |^2],
\label{WF}
\end{eqnarray}
where $\alpha$ and $\beta$ represent complex phase space displacements and $\int \int W_{|\xi\rangle}(\alpha,\beta) ~d^2\alpha d^2\beta=1$, and $\{x,k_x\}$, $\{y,k_y\}$ are conjugate quadrature observables. In terms of the
variables $ X, P_X, Y $ and $ P_Y $, the Wigner function (with the replacements
$\alpha = \frac{X + i P_X}{\sqrt{2}}$, $\beta = \frac{Y + i P_Y}{\sqrt{2}}$, and $\phi=0$) becomes
\begin{eqnarray}
W_{\xi}(X,P_X;Y,P_Y) &=& \frac{1}{\pi^2} \exp[ -2(P_X P_Y-X Y)\sinh 2r \nonumber \\
&&   -(X^2+Y^2+P_X^2 \nonumber \\
&&+P_Y^2)\cosh 2r].
\end{eqnarray}

 Bell violation by the NOPA state has been
studied earlier \cite{PRA'98}. In terms of the Wigner transform $\Pi[\alpha,\beta]$ ($=\frac{\pi^2}{4} W_{|\xi\rangle}[\alpha,\beta]$) the Bell sum is given by \cite{PRA'98}
\begin{eqnarray}
BI&=&\Pi[\alpha=0,\beta=0]+\Pi[\alpha=\sqrt{J},\beta=0]\nonumber \\
&&+\Pi[\alpha=0,\beta=-\sqrt{J}]-\Pi[\alpha=\sqrt{J},\beta=-\sqrt{J}] \nonumber \\
&=& 1+2Exp[-2J \cosh(2r)] \nonumber\\
&& - Exp[-4 J ( \cosh^2(r) - 2 \cos(\phi) \cosh(r) \sinh(r)\nonumber\\
 &&+ \sinh^2(r) ) ], \label{BI_squeezed}
\end{eqnarray}
where $J$ represents amount of displacement in the phase space. By choosing $\phi=0$  and considering $r\rightarrow \infty$ \cite{PRA'98}, the above expression becomes
\begin{eqnarray}
BI(J,r)=1-Exp[-4 J e^{2r}]+2Exp[-J e^{2r}]
\label{BI_squeezed_2}
\end{eqnarray}
The maximum value of $BI$ is $2.19055$ \cite{PRA'98}  (for the above choice of settings) which  occurs for the constraints 
\begin{eqnarray}
J Exp[2 r]=\frac{1}{3} \ln2,
\label{Res}
\end{eqnarray}
where  $J<< 1$.
For example, $BI_{\max}$ ($=2.19055$) occurs for the choice of parameters $J=0.00009467$ and $r=3.9$.  
However,  a more general choice of settings \cite{PRA'03,PRA'04}
\begin{eqnarray}
BI&=&\Pi[\alpha_1,\beta_1]+\Pi[\alpha_1,\beta_2]+\Pi[\alpha_2,\beta_1]\nonumber \\
&& -\Pi[\alpha_2,\beta_2], 
\label{BI_squeezed_M}
\end{eqnarray}
leads to the maximum Bell violation $BI_{\max}=2.32449$ for the choice of 
parameters  $\alpha_1=0.0036990,~\alpha_2=-0.0115244,~\beta_1=-0.0039127,~\beta_2=0.0113108,~r = 3.8853675$.

The subtraction of $n$ photons from the state $|\xi\rangle$ (\ref{NOPA}) may
be represented as
\begin{eqnarray}
|\xi_{n}\rangle=  (a\otimes I + (-1)^k I\otimes b)^n ~ |\xi\rangle,
\label{xi-}
\end{eqnarray} 
where $k\in\{0,1\}$, and it is assumed that one does not know from which
mode the photon is subtracted.  After  normalization the state becomes $\sqrt{N_n} |\xi_{n-}\rangle$, where the normalization constant $N_n$ is given by
$(N_n)^{-1} =\langle\xi_{n}|\xi_{n}\rangle$.
The Wigner function of the state $ |\xi_{n}\rangle $ is related to the Wigner 
function of the state $ |\xi_{(n-1)}\rangle $ by
\begin{eqnarray}
W_{n}(\alpha,\beta) = \hat{\Lambda}(\alpha,\beta)  ~ W_{(n-1)}(\alpha,\beta),
\end{eqnarray}
where the operator $\hat{\Lambda}(\alpha,\beta)$ is given by
\begin{eqnarray}
\hat{\Lambda}(\alpha,\beta) = && [\left(\alpha^\ast + \frac{1}{2} \frac{\partial}{\partial \alpha}\right) \left(\alpha + \frac{1}{2} \frac{\partial}{\partial \alpha^\ast}\right) \nonumber \\
&&+ \left(\alpha^\ast + \frac{1}{2} \frac{\partial}{\partial \alpha}\right) \left(\beta + \frac{1}{2} \frac{\partial}{\partial \beta^\ast}\right) \nonumber \\
&& + \left(\alpha + \frac{1}{2} \frac{\partial}{\partial \alpha^\ast}\right) \left(\beta^\ast + \frac{1}{2} \frac{\partial}{\partial \beta}\right) \nonumber \\
&&+ \left(\beta^\ast + \frac{1}{2} \frac{\partial}{\partial \beta}\right) \left(\beta + \frac{1}{2} \frac{\partial}{\partial \beta^\ast}\right)]. 
\end{eqnarray}
The Wigner function $W_{n}(\alpha,\beta)$ is obtained from $W(\alpha,\beta)$ given by Eq.(\ref{WF}) by applying $\hat{\Lambda}(\alpha,\beta)$ $n$ times, i.e., 
$W_{n}(\alpha,\beta) = \hat{\Lambda}^n(\alpha,\beta)  ~ W(\alpha,\beta)$, and normalizing suitably ($\int W_{n}(\alpha,\beta)~ \mathrm{d}^2\alpha ~ \mathrm{d}^2\beta =1$).
In terms of the $ X, P_X, Y $ and $ P_Y $, the Wigner function for the single
photon subtracted squeezed vacuum state becomes
\begin{eqnarray}
W_{1}(X,Y,P_X,P_Y) &=& \frac{1}{\pi ^2} \exp (2 \sinh (2 r) (X Y-P_X P_Y)\nonumber \\
&&- \cosh (2 r)  (X^2+Y^2+P_X^2+P_Y^2) ) \nonumber \\
&&  (- \sinh (2 r)  (P_X^2-2 P_X P_Y+P_Y^2 \nonumber \\
&&-(X-Y)^2 )+ \cosh (2 r)  (P_X^2 
\label{wigsing} \\
&& -2 P_X P_Y+P_Y^2+(X-Y)^2 )-1 ) \nonumber
\end{eqnarray}

To evaluate the Bell violation, we use the Wigner transform $\Pi_{n}(\alpha,\beta)~(= \frac{\pi^2}{4} W_{n}(\alpha,\beta))$. The Bell sum using the above Wigner transform may be expressed as
\begin{eqnarray}
BI_{n} &=& \Pi_{n}(\alpha_1,\beta_1) + \Pi_{n}(\alpha_1,\beta_2) \nonumber \\
&&+\Pi_{n}(\alpha_2,\beta_1) - \Pi_{n}(\alpha_2,\beta_2)
\label{BI_n-}
\end{eqnarray}
Now, to obtain the maximum Bell violation, one maximizes $BI_{n}$ over $\alpha_1, ~ \alpha_2,~\beta_1,~\beta_2,~r$ for a given value of $n$.

Considering single photon reduction from each mode, i.e., $a\otimes I + (-1)^k I\otimes b$,  the state (\ref{xi-}) becomes
\begin{eqnarray}
|\xi^{1-}\rangle &=& \sqrt{1-\lambda^2} \sum \lambda^n \sqrt{n} [|n-1,n\rangle \nonumber \\
&& + (-1)^k |n,n-1\rangle ]
\label{SPSub}
\end{eqnarray} 
with normalization constant $N_1 = \frac{1}{2\sinh^2(r)}$. The Wigner transform for the above state is given by
\begin{eqnarray}
\Pi_{1}(\alpha,\beta) = && \exp[2 (\alpha \beta +\alpha^\ast \beta^\ast)  \sinh(2r)  - 2  (|\alpha|^2 \nonumber  \\
&&  + |\beta|^2) \cosh(2r)]~(- (2 \alpha \beta + 2 \alpha^\ast\beta^\ast \nonumber \\
&&  + (-1)^k \alpha^2 +(-1)^k (\alpha^\ast)^2 +(-1)^k (\beta^2 \nonumber \\
&& +(\beta^\ast)^2 )) ~  \sinh(2r)   +2   (\alpha  (\alpha^\ast + (-1)^k \beta^\ast) \nonumber \\
&& + \beta  (\beta^\ast + (-1)^k \alpha^\ast ) ) \cosh(2r)-1)  
\end{eqnarray}
The maximum Bell violation, i.e., $(BI_{1})_{\max} =-2.5444$ occurs for the choices $\alpha_1=- 0.0067$, $\alpha_2=0.0201$, $\beta_1=0.0067$, $\beta_2=- 0.0201$, $r=3.0$ and $k=1$. Now, comparing with the two-mode
squeezed state where the Bell violation is $-2.3245$ \cite{paris}, it is seen that by  
photon annihilation, the maximum Bell violation increases.
For the case of two photon subtraction from each mode ($(a\otimes I + (-1)^k I\otimes b)^2$), we can similarly obtain the maximum Bell violation which turns
out to be $(BI_{2})_{\max} =2.6305$  for the choices $\alpha_1=-0.1338$, $\alpha_2=-0.1392$, $\beta_1=-0.1365$, $\beta_2=-0.1311$, $r=4.4015$ and $k=1$. We thus see that the maximum Bell violation increases further.

We have seen in the last section that the Reid criterion is able bring out
the steering property of two mode squeezed vacuum state.
Let us now see whether it is possible to demonstrate steering for single 
photon annihilated state (\ref{SPSub}) using the Reid criterion.
The uncertainty for the inferred observables is in this case given by
\begin{eqnarray}
(\Delta_{\inf} X_{\theta})^2 &=& \cosh(2 r) -\sinh(r) \cosh(r) \cos(2\theta )  \\
&& -\frac{(\cosh(2 r) \cos(\theta -\phi )-2 \sinh(2 r) \cos(\theta +\phi ))^2}{4 (\cosh(2 r)-\sinh(r) \cosh(r) \cos(2 \phi ))}.\nonumber
\end{eqnarray} 
Calculating the minimum value of $(\Delta_{\inf} X_{\theta})^2$ for two different values of $\theta$ (i.e., $\theta_1=0$ and $\theta_2=\pi/2$),  the product of uncertainties turns out to be
\begin{eqnarray}
(\Delta_{\inf} X_{\theta_1})^2 (\Delta_{\inf} X_{\theta_2})^2 = \frac{9}{2 (3 \cosh (4 r)+5)},
\label{SP_NOPA_Reid}
\end{eqnarray}
which goes to $0$ for $r\rightarrow \infty$.  In  the Figure 3a we compare the amount of violation of the Reid inequality by the NOPA and 
the single photon annihilated NOPA states. We see that the Reid criterion fails
in the latter case for smaller values of the squeezing parameter $r$.
 
\begin{figure}[!ht]
\resizebox{9cm}{4cm}{\includegraphics{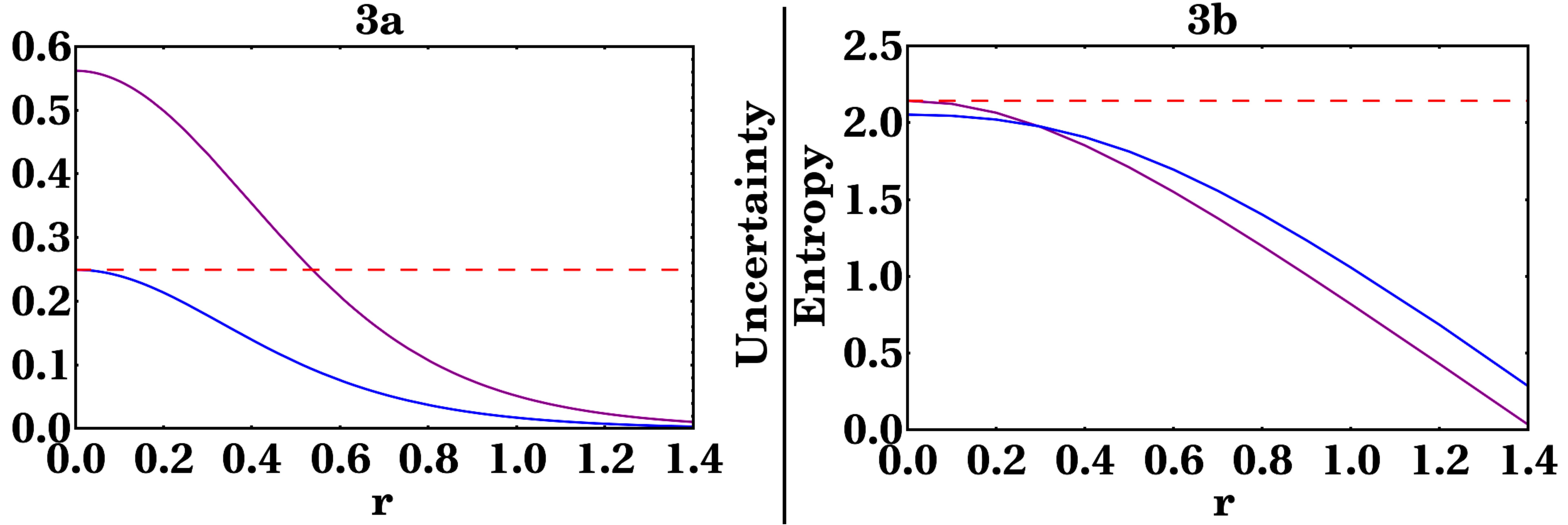}} 
\caption{\footnotesize (Coloronline)
\textit{a} : The horizontal line represents the uncertainty bound below which
steering is signified. The lower curve represents the product of inferred uncertainties for the two-mode
squeezed vacuum state. Steering is demonstrated for all values of $r$ through 
the Reid criterion. The upper curve represents the product of uncertainties 
for the photon subtracted state. Clearly, the Reid criterion fails to
show steering for smaller values of $r$ in the latter case.\\
\textit{b} : The horizontal line represents the bound $ \ln \pi e $. The purple and
blue  curves represent the LHS of the steering inequality  for the squeezed
 state and the  single photon subtracted state,  respectively.
}
\label{Fig_Reid}
\end{figure}
We next demonstrate steering for the photon subtracted squeezed vacuum state
through the entropic steering inequality. Considering the measurements 
corresponding to either position ($ r=x $) or momentum ($ s=p $), for 
the single photon subtracted squeezed vacuum state correlations exist between 
$ X $ and $ Y $, and $ P_X $ and $ P_Y $.  $\{X,~P_X\}$ and $\{Y,~P_Y\}$ are conjugate pairs of dimensionless quadratures. So in terms of these variables,
the steering inequality (\ref{entropy_steering}) becomes
\begin{eqnarray}
h(\mathcal{Y}|\mathcal{X})+h(\mathcal{P_Y|}\mathcal{P_X})\geq \ln \pi e,
\label{NOPA_entropy_steering}
\end{eqnarray}
where $ X,~Y,~P_X, $ and $ P_Y $ are the outcomes of measurements $ \mathcal{X},~\mathcal{Y},~\mathcal{P_X}, $ and $ \mathcal{P_Y} $ respectively.
Here, the conditional entropies $ h(\mathcal{Y}|\mathcal{X}) $ and $ h(\mathcal{P_Y}|\mathcal{P_X}) $ are given by
\begin{eqnarray}
h(\mathcal{Y}|\mathcal{X})&=& h(\mathcal{X},\mathcal{Y})-h(\mathcal{X}),\nonumber\\
h(\mathcal{P_Y}|\mathcal{P_X})&=&h(\mathcal{P_X},\mathcal{P_Y})-h(\mathcal{P_X}),
\label{condt_prob}
\end{eqnarray}
and  calculated using the marginal probability distributions obtained 
from the Wigner function (\ref{wigsing}). 
One can thus calculate the L.H.S. of the inequality (\ref{NOPA_entropy_steering}) 
for the single photon subtracted state for any value of the squeezing
parameter $r$. In Fig. 3b we plot the L.H.S. of 
the entropic steering inequality versus $ r $ for the  squeezed vacuum state
as well as the single photon subtracted state. The figure shows the violation 
of the steering inequality  increases with $ r $ for each of these two states.

In the following table we show the comparison of Bell violation with  entropic EPR steering for the NOPA state and the single photon annihilated NOPA state.
Note here that $\frac{|BI_{\max}|}{2} > 1$ signifies Bell violation, and
$\frac{(\ln\pi e)}{h(\mathcal{X}|\mathcal{P_Y})+h(\mathcal{P_X|}\mathcal{Y})} >1$ identifies steering. One sees that though the magnitude of Bell violation reaches a maximum
for a certain value of the squeezing parameter $r$, and subsequently decreases
gradually, the strength of steering increases monotonically 
with $r$. Hence, it would be much easier to observe steering compared to
Bell violation for higher values of $r$.

\begin{tabular}{|c|c|c|c|}
\hline 
State & $r$ & Bell violation   & Entropic EPR steering criterion\\ 
      &     &   ($=\frac{|BI_{\max}|}{2}$)   & $(=\frac{\ln(\pi e)}{h(\mathcal{Y}|\mathcal{X})+h(\mathcal{P_Y|}\mathcal{P_X})})$\\
\hline 
$|\xi\rangle$ & 0 & 1.0 & 1.0 \\ 
\hline 
$|\xi\rangle$ & 0.2 & 1.040 & 1.038 \\ 
\hline
$|\xi\rangle$ & 0.4 & 1.091 & 1.157 \\ 
\hline
$|\xi\rangle$ & 0.6 & 1.125 & 1.383 \\ 
\hline
$|\xi\rangle$ & 0.8 & 1.144 & 1.790 \\ 
\hline 
$|\xi\rangle$ & 1 & 1.153 & 2.616 \\ 
\hline 
$|\xi\rangle$ & 1.2 & 1.159 & 4.991 \\ 
\hline 
$|\xi\rangle$ & 1.4 & 1.160 & 62.737 \\ 
\hline
\hline
$|\xi_{1}\rangle$ & 0 & 1.120 & 1.044 \\ 
\hline
$|\xi_{1}\rangle$ & 0.2 & 1.189 & 1.061 \\ 
\hline
$|\xi_{1}\rangle$ & 0.4 & 1.229 & 1.124 \\ 
\hline
$|\xi_{1}\rangle$ & 0.6 & 1.252 & 1.264 \\ 
\hline
$|\xi_{1}\rangle$ & 0.8 & 1.263 & 1.529 \\ 
\hline
$|\xi_{1}\rangle$ & 1 & 1.267 &  2.027 \\ 
\hline
$|\xi_{1}\rangle$ & 1.2 & 1.271   & 3.132 \\ 
\hline
$|\xi_{1}\rangle$ & 1.4 & 1.271  & 7.531 \\ 
\hline
\end{tabular}

\subsection{N00N state}

The maximally path-entangled number states have the form given by
\begin{eqnarray}
|\psi\rangle = \frac{1}{\sqrt{2}}(|N\rangle_{a}|0\rangle_{b}+e^{i \phi}|0\rangle_{a}|N\rangle_{b}).
\label{N00N state}
\end{eqnarray}
This is an example of a two-mode state such that $ N $ photons can be found either in the mode $a$ or in the mode $b$, and is referred to as `$ N00N $' states
\cite{dow}.
The utility of N00N states in making precise interferometric measurements
is of much importance in quantum metrology. Such states have been recently
experimentally realized up to $N=5$ \cite{noonexpt}. The entanglement
of N00N states is obtained in terms of the logarithmic negativity, {\it viz.}
$E_N = 1$ \cite{book}, a value that is independent of $N$.

The Wigner distribution function for the $ N00N $ state is given by \cite{Bell_N00N}
\begin{eqnarray}
W(\alpha,\beta) &=& \frac{2}{\pi^2} e^{-2|\alpha|^2-2|\beta|^2}[(-1)^N(L_N(4|\alpha|^2)+L_N(4|\beta|^2)) \nonumber \\
&& -\frac{2^{2N}}{N!}(\alpha^{*N} \beta^N+\alpha^N \beta^{*N})],
\end{eqnarray}
where for simplicity we choose $\phi=\pi$ and $ L_N(x) $ is the Laguerre polynomial.
In terms of the dimensionless quadratures $ \{X, P_X\} $ and $ \{Y, P_Y\} $ 
the Wigner function becomes
\begin{eqnarray}
W(X, P_X, Y, P_Y) =&& \frac{1}{2\pi ^2 N!} e^{-(X^2+Y^2+P_X^2+P_Y^2)}\nonumber\\
&& [ -2^N \{(X+i P_X)^N (Y-i P_Y)^N \nonumber \\
&& +(X-i P_X)^N (Y+i P_Y)^N\}+\nonumber\\
&&(-1)^N N! \{L_N\left(2(X^2+P_X^2)\right)\nonumber \\
&& +L_N\left(2(Y^2+P_Y^2)\right)\}]   .
\label{Wigner_X_Y}
\end{eqnarray}
The Bell-CHSH inequality 
\begin{eqnarray}
|BI| = \Pi(\alpha, \beta)+\Pi(\alpha^{\prime}, \beta)+\Pi(\alpha, \beta^{\prime})-\Pi(\alpha^{\prime}, \beta^{\prime}) \leq 2
\label{BCHSH}
\end{eqnarray}
is maximally violated with $BI_{\max} = -2.2387 $ which occurs for $ N=1 $ and the corresponding settings are $\alpha=- \beta=0.0610285$, $\alpha^{\prime}=-\beta^{\prime}=-0.339053$.  States with larger $ N $ do not violate the inequality.
However, there are some other Bell-type inequalities \cite{Bell_N00N} for 
six correlated events for which $ N00N $ states show the violation for 
any $ N $.

From the expression of the Wigner function (\ref{Wigner_X_Y}) for the 
N$00$N states the presence of correlations of the type $\langle X,Y\rangle$ is
clear. Using such correlations
the entropic steering inequality for the N00N state may be written as
\begin{eqnarray}
h(\mathcal{Y}|\mathcal{X})+h(\mathcal{P_Y|}\mathcal{P_X})\geq \ln \pi e,
\label{N00N_entropy_steering}
\end{eqnarray}
The conditional entropies $ h(\mathcal{Y}|\mathcal{X}) $ and $ h(\mathcal{P_Y|}\mathcal{P_X}) $ can be calculated through the marginal probabilities
obtained through the Wigner function (\ref{Wigner_X_Y}), using which the 
L.H.S. of the inequality (\ref{N00N_entropy_steering}) may be obtained for 
different values 
of $ N $. It turns out that for $N=1$, one gets $h(\mathcal{Y}|\mathcal{X})+h(\mathcal{P_Y|}\mathcal{P_X}) \approx 2.05 < \ln \pi e$, thus violating the
steering inequality. However, for $N=2$, one gets  $h(\mathcal{Y}|\mathcal{X})+h(\mathcal{P_Y|}\mathcal{P_X}) \approx 2.25 > \ln \pi e$. Larger values of $N$
lead to further higher values of $h(\mathcal{Y}|\mathcal{X})+h(\mathcal{P_Y|}\mathcal{P_X})$, and hence, no steering
is possible for $N>1$. 

\begin{figure}[!ht]
\resizebox{9cm}{6cm}{\includegraphics{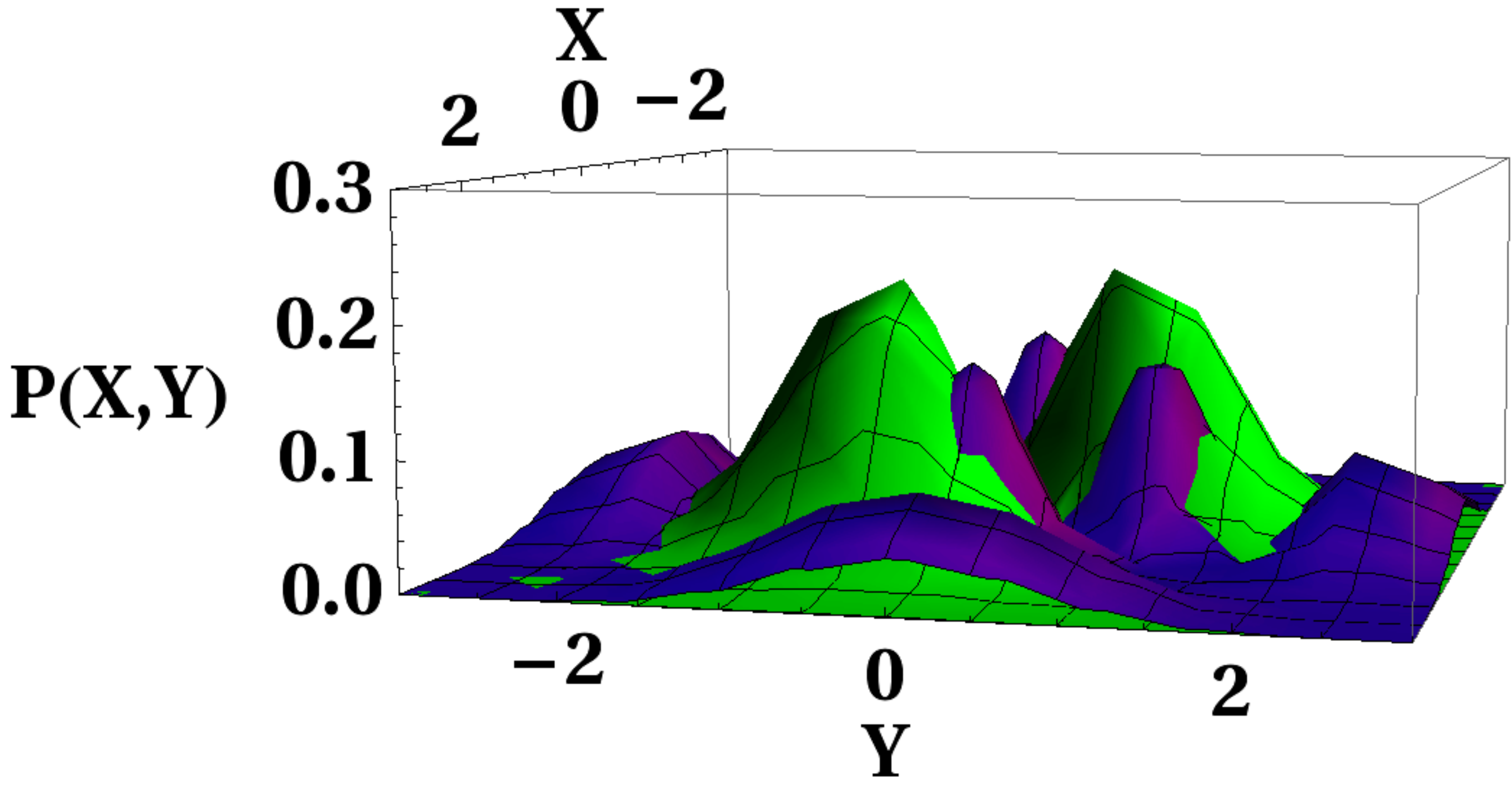}} 
\caption{\footnotesize (Coloronline)
Correlations of the type $\langle X,Y\rangle$ responsible for steering using the
entropic steering inequality are revealed through the joint probability
distributions $P(X,Y)$. The figure shows that such correlations are 
sufficiently strong to admit steering
for $N=1$, but are significantly weakened for larger $N$.
}
\label{PXY_N_N00N_F}
\end{figure}

In Fig.[\ref{PXY_N_N00N_F}], we plot the joint probability $P(X,Y)$ for
two different values of $N$, {\it viz}., $N=1$ and $N=4$, respectively.
The higher peak of the $N=1$ curve indicates stronger $\langle X,Y\rangle$ correlations
responsible for steering in this case. The correlations weaken for larger
values of $N$ as is indicated by the lower peak value of the $N=4$ curve,
and are not sufficient for revealing steering through the entropic inequality.  
Thus,  $ N00N $ states 
with  $ N=1 $ 
violate the entropic steering inequality, but for $ N\geq 1 $,  these states are
not steerable. This feature is similar to Bell violation for $N00N$ states
which is revealed for $N=1$, but the violation of the standard Bell-CHSH 
inequality does not occur for $N \ge 1$.

\section{Summary}

In the present paper we have studied EPR steering by non-Gaussian
continuous variable entangled states. Here we have considered several
examples of such systems, i.e., the two-dimensional harmonic oscillator,
the photon subtracted squeezed vacuum state, and the N00N state.
Though such states are entangled
pure states, we have shown that they fail to reveal steering through
the Reid criterion for wide ranges of parameters. Steering with
such states is demonstrated using the entropic steering inequality.
We have computed the relevant conditional entropies using the
Wigner function whose non-Gaussian nature plays an inportant role
in demonstrating steering.
For all the above examples we perform a quantitative study of the
strength of steering (determined by the magnitude of violation of
the entropic steering inequality) as a function of the state parameters.
This leads to some interesting observations, especially in comparison
with the magnitude of Bell nonlocality demonstrated by these states.

For the LG modes one sees that the steering strength increases with
the increase of the angular momentum $n$, a feature that is also common
to the Bell violation. However, for both the two-mode squeezed  vacuum
state as well as the single photon subtracted state derived from it,
we show that the behavior of the maximum Bell violation and steering
strength versus the squeezing parameter are not similar. This is evident
from the fact that though the maximum Bell violation peaks for a certain
value of $r$, the steering strength rises monotonically with increasing
$r$. This feature clearly establishes the fact that though Bell violation
guarantees steerability, the two types of quantum correlations are distinct
from each other. Moreover, the presence of quantum correlations in such
class of states may be more easily detected through the violation of the 
entropic steering inequality compared to the violation of the Bell inequality
for higher values of squeezing.
Finally, we study steering by N00N states. Here, steering
through the entropic steering condition is revealed only for  $N=1$, though
the entanglement of such states remains constant with $N$. This shows
that entanglement is a different correlation compared to steering, 
as also it is different compared to Bell
nonlocality. The above results should be useful for detecting and 
manipulating correlations in non-Gaussian states for practical purposes
in different arenas such as information processing, quantum metrology, and
Bose condensates. Further work on the issue of the recently proposed
symmetric steering framework \cite{schn} may be of interest using
non-Gaussian resources.

{\it Acknowledgements}: GSA thanks Tata Institute of Fundamental Research 
where a part of this work was done.

\end{document}